\documentstyle[12pt]{article}

\newcommand{\eq}[1]{eq.(\ref{#1})}

\def\be{\begin{equation}}
\def\ee{\end{equation}}

\def\TeV{\mbox{TeV}}

\def\log{\mbox{log}}

\def\L{\Lambda}
\def\expansion{$(2+\epsilon)$ expansion }

\def\function{$\beta$-function }
\def\NG{Nambu-Goldstone }
\begin{document}
\begin{flushright}
UCLA/93/TEP/29 \\
September 1993 \\
hep-ph/9310221
\end{flushright}
\vspace{1in}
\begin{center}
{\Large Strongly interacting $W$ bosons and supersymmetry} \\
\vspace{0.4in}
{\large S.Yu.~Khlebnikov}
\footnote{On leave of absence
from Institute for Nuclear Research of the Academy
of Sciences, Moscow 117312 Russia.
Address after 1 December 1993: Department of Physics, Purdue University,
West Lafayette, IN 47907} \\
\vspace{0.2in}
{\it Department of Physics, University of California,
Los Angeles, CA 90024, USA} \\
\vspace{0.7in}
{\bf Abstract} \\
\end{center}
We present arguments in favor of the idea that supersymmetric sigma models
with compact symmetric K\"ahler spaces as target manifolds have a
second-order phase transition in four dimensions. When applied to electroweak
symmetry breaking, these models then do not require a light Higgs
boson or techni-resonances but predict fermionic superpartners of longitudinal
W's and Z with masses at or below TeV scale. Presence of the phase transition
leads to a fractional-power energy dependence of fixed-angle scattering
amplitude of longitudinal W's and Z's at high energies.
\newpage
The existing ways of electroweak symmetry breaking require either
a light, subject to triviality \cite{triviality} bound, Higgs boson
or, as in technicolor \cite{techni}, other bosonic resonances at
approximately 1 TeV. In this note we propose to consider an alternative --
situation when a four-dimensional theory has a second-order phase
transition. We are interested in cases when, with respect to some effective
temperature parameter, a low-temperature phase describes interacting
\NG bosons in four dimensions. At some critical temperature there is a phase
transition with non-mean-field values of critical exponents.
We stress that that the word 'temperature' refers essentially to some
dimensionless coupling constant. At any value of this constant, the
system remains Lorentz-invariant in the four-dimensional sense. Hence,
'temperature' in our meaning of the word has nothing to do with
actual heating of the system.

Phase transitions of the above type are not commonly known in four
dimensions and it is easy to see why. Interacting \NG bosons are
described by non-linear sigma models and, in four dimensions, these
are usually trivial because of the equivalence with linear ones.
Indeed, the pattern of \NG bosons in a non-linear
sigma model can usually be obtained via spontaneous symmetry breaking
by elementary Higgs boson(s) in a linear sigma model.
Still, there are cases when such equivalence (involving all massless
particles, both bosons and fermions) may not be possible.
These cases are {\it supersymmetric}
non-linear sigma models with homogeneous K\"ahler spaces as target
manifolds. It is known \cite{doubling} that the pattern of \NG bosons
of these models cannot be produced by spontaneous breaking of a
global symmetry by elementary Higgs bosons in any linear
supersymmetric sigma model.
In two and three dimensions, the equivalence can still be achieved by
introduction of an abelian gauge field with a Fayet-Iliopoulos term
in action \cite{equiv}.
In four dimensions, however, chiral fermions prevent introduction
of the required gauge field because of anomaly.

Our argument in favor of a phase transition in supersymmetric
non-linear sigma models, for a certain class of homogeneous K\"ahler
spaces, is based on renormalization group analysis. In non-linear sigma
models we consider below,
there is a dimensionful coupling constant, which is analogous to
$f_{\pi}$ of pion physics and which we denote by $F$. Dimensionless
effective temperature $t$ is constructed from the coupling $F$ taken at a
normalization scale $\mu$ and the scale $\mu$ itself. In $d$ dimensions,
we define
\be
t(\mu) = \frac{S_d \mu^2}{(2\pi)^d F^2(\mu)} \; ,
\label{t}
\ee
where $S_d$ is the area of unit sphere. Slightly above two dimensions,
the \function for the coupling $t$ may be found by \expansion
\cite{exp1,exp2}.
An invariant regularization in our case is provided by dimensional
reduction \cite{reduction}. To one-loop order,
\be
\beta(t) = (d-2) t - b t^2 \; ,
\label{beta}
\ee
where for {\em compact} models $b>0$. Hence, there is an ultraviolet
stable fixed point for $t$ at $t=t_c=(d-2)/b$ indicating a second-order
phase transition. In the special case of
supersymmetric non-linear sigma models with {\em symmetric}
compact K\"ahler spaces as target manifolds, there are no higher-loop
corrections to the two-dimensional $\beta$-function.
This was argued via different lines in refs.\cite{arg1,arg2} and verified
up to the four-loop order by explicit calculation in ref.\cite{explicit}.
For these models, \eq{beta} does not acquire terms of higher order in $t$
and may be continued all the way to four dimensions.
This suggests that a second-order phase transition exists in these
models even in $d=4$.

We will shortly reformulate this argument directly from the four-dimensional
point of view. The reason why this argument is still not
sufficient to firmly establish the existence of a phase transition
is that in sigma models in four dimensions the usual perturbative
expansion produces divergences that require higher-derivative operators
as counterterms. This fact is known in the context
of chiral perturbation theory \cite{chiral}. We cannot argue
that these higher-derivative operators are irrelevant in the
renormalization group sense and, in fact, this may well not be true --
for a recent discussion in the context of bosonic sigma models
see ref.\cite{rel}. However, if we first
{\em assume} that the existence of a fixed point for $t$ is equivalent
to the existence of a phase transition, then we can put forward (see below)
a resummation procedure that eliminates divergences in calculations of
physical quantities (such as scattering amplitudes) without the use of
higher-derivative counterterms. This makes the assumption of a phase
transition at least self-consistent.

Let us first consider renormalization of $F$ in more detail.
In four dimensions, the corresponding ultraviolet divergences are
quadratic and can in principle be subtracted at zero momenta by using,
say, a version of minimal subtraction scheme. For our purposes,
however, it is essential to be able to subtract these divergences at
arbitrary renormalization scale $\mu$. This will allow us to consider
changes in $F(\mu)$ as $\mu$ is varied and thus apply
the renormalization group to establish the existence of a phase transition.
The only quadratic divergence of supersymmetric sigma models with
symmetric K\"ahler target spaces is the one-loop tadpole.
The corresponding renormalization of $F$
is best formulated in the background field method \cite{background}.
If we denote the bare coupling by $F_B$, the renormalized coupling
at scale $\mu$ is given schematically by
\be
F^2(\mu) = F^2_B - {\tilde b} \int_{|p|>\mu} \frac{d^4 p}{(2\pi)^4}
\frac{1}{p^2} \; ,
\label{schem}
\ee
where ${\tilde b}>0$ and only momenta bigger than $\mu$ are taken into
account in the tadpole integral. Sharp momentum cutoff is indicated
in (\ref{schem}) only for illustrative purposes, we actually need a
procedure that preserves invariances of the theory.
Such a procedure can be formulated as follows.

We start with the theory defined in $(2-\delta)$ dimensions and then
analytically continue to $d=4$. This allows one to use two-dimensional
$N=1$ superfield language compatible with the normal coordinate version
of the background field expansion (cf. ref.\cite{explicit}).
When restricted to K\"ahler geometry and analytically
continued to $d=4$, the result will produce counterterms of
the four-dimensional sigma model. The normal coordinate superfields may
be expanded with respect to a tangent frame of the sigma-model manifold.
The superfields that are coefficients of this expansion transform as
vectors with respect to tangent frame rotations and are scalars under
reparametrizations of the manifold. Hence, one may introduce a mass
for these superfields without breaking any invariances of the model.
This leads to the following replacement of the non-invariant \eq{schem}
\be
F^2(\mu) =
F^2_B + {\tilde b} \int\frac{d^d p}{(2\pi)^d}\frac{1}{p^2+2\mu^2} -
2{\tilde b} \int\frac{d^d p}{(2\pi)^d}\frac{1}{p^2+\mu^2} \; .
\label{subtr}
\ee
Other choices of the ratio of two masses in (\ref{subtr}) are possible,
they lead to different values of $\mu$ dependent terms and, hence,
of critical coupling but to the same value of the critical exponent
$\nu'$.

A straightforward calculation of the integrals in (\ref{subtr})
via dimensional regularization gives for $d=4$
\be
F^2(\mu) = F^2_B + \frac{{\tilde b} \log 2}{8\pi^2} \mu^2 \; .
\label{fmu}
\ee
Hence, in dimensional regularization $F^2_B$ is actually $F^2(0)$.
As we already noted, the coefficient of the $\mu^2$ term is non-universal.
Eq.(\ref{fmu}) shows the existence of an ultraviolet stable fixed point
for the dimensionless coupling $t$ at $t=t_c=({\tilde b} \log 2)^{-1}$.
Eq.(\ref{fmu}) can be rewritten as follows,
\be
8\pi^2 F^2(0) = \mu^2 \left( \frac{1}{t(\mu)} - \frac{1}{t_c} \right) \; .
\label{scal}
\ee
When existence of a fixed point for $t$ is interpreted as existence
of a phase transition, renorm-invariant quantity
\be
\L = 2\sqrt{2} \pi F(0) = \mu \left( \frac{1}{t(\mu)} - \frac{1}{t_c}
\right)^{\nu'}
\label{L}
\ee
is the inverse of a correlation length and the corresponding critical
exponent is $\nu'=1/2$. For the electroweak theory, $\L\sim 2~\TeV$.

The physical meaning of correlation length in a low-temperature phase
\cite{Josephson,exp2} is that it determines a scale
at which the full propagator of a \NG boson (and, in our case, also of its
superpartner) changes behavior from low-momentum $1/p^2$ to critical
$1/p^{2-\eta}$, $\eta>0$. Note that in our
case there are no higher-loop corrections to the exponent $\nu'$, so that
the value $\nu'=1/2$ (in $d$ dimensions, $\nu'=1/(d-2)$) is exact.
This does not mean by itself that the phase transition in four dimensions
is of the mean-field type - we see no reason why in our case $\eta$ should
vanish. The scale $\L$ is of the same order
as the energy scale at which tree-level partial-wave unitarity in scattering
of two \NG bosons breaks down \cite{unitarity}. When there is a second-order
phase transition, unitarity above this scale is restored via quantum
effects in interactions of \NG bosons (and, in our case, their superpartners)
without a light Higgs boson or other resonances needed. Instead, our
theory predicts fermionic superpartners of longitudinal W's and Z,
with masses at or below TeV scale. Besides, since the number of \NG
bosons in a K\"ahler sigma model is even, there is at least one
(pseudo)\NG boson in addition to those that become longitudinal
components of W and Z.

The resummation procedure that removes the need for higher-dimensional
counterterms is based precisely on the property that,
under the assumption of a second-order phase transition,
momentum dependence of full propagators changes to $1/p^{2-\eta}$
at large Euclidean momenta. From now on, we choose normalization point for $F$
at $\mu=0$. This allows to understand all loop integrals in the sense of
analytic continuations from lower dimensions, without adding or subtracting
anything. Then, using full propagators instead of bare ones in internal lines
of diagrams makes the corresponding integrals convergent in this sense
whenever $\eta$ is irrational.
Consider for example fixed-angle two-into-two scattering of \NG bosons
(which, by the equivalence theorem \cite{equiv1,unitarity,equiv2}, also gives
scattering of longitudinal W's) as a function of center-of-mass energy $E$.
The most interesting region is that of high energies $E\gg \L$.
The tree amplitude is of order $E^2/F^2(0)$. At one-loop level,
for $E\gg \L$, the main contribution comes from internal momenta
in the critical region. The amplitude is
\be
A(E) \sim \frac{E^2}{F^2(0)} + c(\eta)\Gamma(-\eta)
\frac{E^{4+2\eta}}{F^{4+2\eta}(0)} + ... \; ,
\label{A}
\ee
where $c(\eta)$ is some $\eta$-dependent constant and we write separately
the factor of $\Gamma(-\eta)$ that produces $1/\eta$ singularity when
$\eta$ becomes small. Dots stand for higher-loop terms. We have used the
fact that $\L\propto F(0)$.
A summation of infinite series is needed to make the amplitude (\ref{A})
explicitly unitary. A large-$N$-type summation gives
\be
A(E) \sim \frac{E^2}{F^2(0)}
\left( 1- c(\eta)\Gamma(-\eta) \frac{E^{2+2\eta}}{F^{2+2\eta}(0)}
\right)^{-1} \sim \frac{1}{\Gamma(-\eta)} \frac{\L^{2\eta}}{E^{2\eta}} \; .
\label{sum}
\ee
Eq.(\ref{sum}) makes clear that it is $\eta$ that determines deviations
from triviality -- at $\eta\to 0$ the amplitude (\ref{sum}) would vanish.

To summarize, we have presented arguments in favor of the idea that
supersymmetric sigma models with compact symmetric K\"ahler
spaces as target manifolds have a second-order phase transition
in four dimensions. When applied to electroweak symmetry breaking,
these models then do not
require a light Higgs boson or techni-resonances but predict fermionic
superpartners of longitudinal W's and Z with masses at or below TeV scale.
Presence of the phase transition leads to a fractional-power energy
dependence of fixed-angle scattering amplitude of longitudinal W's and
Z's at high energies.

The author is grateful to M. Bos, S. Chakravarty, M. von Ins, S. Love,
D. Morris and R. Peccei for very useful discussions. The author is
supported by the Julian Schwinger fellowship at UCLA.

\end{document}